\documentstyle{mn}

\newcommand{\etal}{\mbox{et\ al.\ }} 
\newcommand{\eg}{\mbox{e.g.\ }} 
\newcommand{\ergsec}{\,\mbox{$\mbox{erg}\,\mbox{s}^{-1}$}}

\def\spose#1{\hbox to 0pt{#1\hss}} 
\def\simlt{\mathrel{\spose{\lower 3pt\hbox{$\mathchar"218$}}
     \raise 2.0pt\hbox{$\mathchar"13C$}}}
\def\simgt{\mathrel{\spose{\lower 3pt\hbox{$\mathchar"218$}}
     \raise 2.0pt\hbox{$\mathchar"13E$}}}
 
\title[Optical Identification of RX~J0059.2-7138]{The Optical Counterpart of 
the Supersoft Small Magellanic Cloud Transient Pulsar RX~J0059.2-7138} 

\author[K.\, A.\ Southwell \& P.\,A.\ Charles]
{K.\, A.\ Southwell and P.\, A.\ Charles \\ 
Department of Astrophysics, Nuclear Physics Building, Keble Road, Oxford~OX1\,
3RH\\}
\date{}

\begin{document}
\maketitle

\begin{abstract}

We identify the probable optical counterpart of the SMC supersoft source, 
RX~J0059.2-7138, with a $\sim$~14th magnitude star lying within 
the X-ray error circle. We present high resolution optical 
spectroscopy, which  
reveals strong HI emission and HeI absorption, indicative 
of a Be star. This classification is consistent with the colours we 
derive from optical photometry. 
We thus find evidence to support the existing tentative identification 
of this object as a 
Be/X-ray binary, the first of its kind to exhibit luminous supersoft X-ray 
emission. 

\end{abstract}

\begin{keywords}
stars: emission line, Be -- pulsars: individual: RX~J0059.2-71 -- 
accretion, accretion discs -- binaries: spectroscopic -- X-rays: stars. 
\end{keywords}

\section{Introduction}

The supersoft X-ray sources (SSSs) have been established by {\it ROSAT} 
observations 
as a new class of objects, characterised by their luminous 
($L_{\rm bol} \sim~10^{37} - 10^{38}$\ergsec) emission at very soft X-ray 
energies (T$_{\rm bb} \sim 30-50$~eV) - see reviews by \eg 
Kahabka \& Tr\"{u}mper (1996); Cowley \etal (1996). The group as a whole is 
not strictly homogeneous; the archetypal sources, CAL~83 and CAL~87, originally 
discovered by the {\it Einstein} X-ray Observatory (Long, Helfand \& 
Grabelsky, 1981) have the highest luminosities, and are popularly 
thought to be white dwarf binaries accreting at exceptionally high 
rates, due to thermally unstable mass transfer from a more massive donor star 
(van den Heuvel \etal 1992). However, luminous supersoft emission has also 
been observed from symbiotic systems (\eg Hasinger 1994) and a planetary 
nebula nucleus (Wang 1991), although their bolometric luminosities are 
typically an order of magnitude lower. 

The supersoft source RX~J0059.2-7138 (hereafter RX~J0059-71)  
was detected serendipitously with the {\it ROSAT} PSPC in 1993 May, at 
a count rate of $\sim 8$\,s$^{-1}$, and was seen almost simultaneously 
by {\it ASCA} (Hughes 1994; Kylafis 1996). 
Previously, it had failed to be detected by 
either the {\it Einstein} Observatory or {\it EXOSAT} in the early 1980s, or in 
recent pointed {\it ROSAT} observations of 1991 November (see Hughes 1994). The 
transient nature of this source is thus clearly established. 

The best fit to 
the X-ray spectrum consists of three components (Kylafis 1996): 
two power laws with indices of 0.7 and 2.0 fit the spectrum in the 
$>3$~keV and $0.5-3.0$~keV bands respectively. Furthermore, the emission 
is pulsed at a level of $\sim 35 \%$ and $\sim 20 \%$ in these respective 
bands, with a period of $\sim 2.7$~s (Hughes 1994), 
firmly establishing the presence of a neutron star. 
{\it However, the lower energy band ($<0.5$\,keV) requires the 
third spectral fit 
component, which accounts for $\sim 90 \%$ of the bolometric luminosity, and 
is a black body with a temperature $kT_{\rm bb} \sim 35$~eV}. 
Such luminous supersoft emission is clearly unexpected in this object, given 
the van den Heuvel \etal (1992) accreting white dwarf binary model for the 
SSSs. 

On the basis of its X-ray properties, Hughes (1994) tentatively identified 
RX~J0059-71 as a high mass X-ray binary, possibly a Be/X-ray binary system 
(see \eg Parmar 1994 for a review of these objects). We have therefore 
undertaken optical spectroscopy and photometry of the proposed counterpart, 
to investigate the nature of this potentially unusual member of the SSSs. 

\section{Optical Spectroscopy and Photometry}

\subsection{Observations and Reduction}

Spectroscopy of the proposed counterpart was obtained in Nov/Dec 1994 
using the 3.9\,m Anglo-Australian Telescope at Siding Spring. The detector 
was a 1024$\times1024$ TEK CCD attached to the RGO spectrograph. The 1200V 
grating on the 25\,cm camera 
gave a resolution of $\sim1.3$\,\AA. In addition, a single spectrum was 
obtained using an identical set-up in Jan~1996. 
Tab.~1 lists the journal 
of spectroscopic observations. Cu-Ar arc spectra were taken 
before or after each 
object exposure, and a series of tungsten lamp flat fields and bias frames were 
also obtained. Owing to poor observing conditions (variable cloud) during both 
runs, no flux standards were 
observed, and hence the data were not corrected for instrumental response.  
Reduction of the data frames followed standard procedures. 
We subtracted the bias signal, and removed small scale pixel-to-pixel 
sensitivity variations by multiplying by a balance frame prepared from the 
tungsten lamp flat fields.  
One dimensional spectra were extracted using the
optimal algorithm of Horne (1986), and conversion of the pixel scale to 
wavelength units was achieved using the arc calibration spectra. 

CCD photometry of the target field 
was obtained in 1996 Jan, using the UCT CCD on the 1.9m telescope 
at the South African Astronomical Observatory, Sutherland. We used 
standard `BV (Johnson) and R (Cousins) filters. Flat fields of the 
twilight sky, and observations of Magellanic Cloud standard stars 
were obtained in each filter. 
The images were reduced using {\sc daophot} (Stetson 1987), since we 
found a profile-fitting technique to be essential 
given the particularly poor seeing ($\sim 2-3$'') during most of the run 
(see Sec.~3.2). 
We show in Fig.~1 an $I$ band finding chart for the target (Star\,1), and 
the star 
relative to which we performed differential photometry (Star\,2) - see 
Sec.~3.2.

\section{Results}

\subsection{Average spectra}

We show in Fig.~2 the variance-weighted average of our four 1994 blue 
spectra; the continuum has been normalised, but the spectra are 
not flux calibrated. 
He{\sc i} absorption is highly prominent, being 
observed at $\lambda\lambda$4388, 4471, 4713, 4922, 5016 and 5048. 
We see also weaker absorption lines of 
of Si{\sc iii} $\lambda\lambda$4553,4568,4575 and 
O{\sc ii} $\lambda$4415-17 and $\lambda$4639-42. The absorption at  
$\lambda \sim$4650 is ambiguous, possibly arising from 
O{\sc ii} $\lambda$4650 or the 
C{\sc iii} $\lambda$4647-51 absorption blend. 

In emission, H$\beta$ appears very 
strongly, but with broad underlying absorption. There is evidence for 
Fe{\sc ii} $\lambda$5018 in emission; however, if this is present, we 
might also expect to see features due to this ion at $\lambda\lambda$4489-91, 
4508-23, 4584 and 4629. There is 
no obvious emission at these wavelengths (save perhaps at $\lambda$4584), 
but a higher signal-to-noise ratio spectrum would be required to 
conclusively confirm the presence of this species. 

The single blue observation obtained in Jan~1996 
suffers from very poor signal-to-noise and is not shown. 
However, we do still see H$\beta$ in emission, and some of the stronger 
He{\sc i} absorption lines, consistent with the spectra obtained $\sim 1$~ 
year earlier. 

In Fig.~3, the variance-weighted average of the two 1994 red spectra is 
presented. H$\alpha$ appears very strongly in 
emission, consistent with CTIO Schmidt narrow-band filter imaging data 
obtained in 1993~Dec (Hughes, private communication). 
We do not see evidence for any intrinsic structure in this line. 
The absorption at $\sim 6270-6295$~\AA\ is an atmospheric feature, but there 
are no other obvious spectral features.

\subsection{Optical Photometry}

Differential photometry was performed relative to 
Star\,2 (see Fig.~1). We used observations of a standard star from the only 
photometric night (1996~Jan\,28) to calibrate the local standard, for which we 
measure $V=16.51$, $B-V=0.82$ and $V-R=0.49$. 
However, we estimate that a systematic error of $\simlt 0.05$~mags is possible, 
given the paucity of photometric standard star observations. 

We note that RX~J0059-71 has a close companion about 
4'' to the south-east (Star\,3 of Fig.~1). Although these two stars were not 
fully resolved on all our data frames, we were able to subtract the 
contaminant star through a profile fitting technique, in the $V$ and $R$ bands. 
The magnitudes of Star\,1 thus derived were consistent with those obtained from 
the fully resolved frames, hence we believe the deblending procedure to be 
effective. We estimate that Star\,3 has $V\sim 16.8$, and $V-R \sim\,0.6$; 
however, we were unable to obtain an accurate measurement of its $B$ magnitude, 
since it appeared almost totally blended with Star\,1 on all frames. 

We list in Tab.~2 the nightly colours of Star\,1 for the period 
1996\,Jan~24-29 (blended with the light of Star\,3 in the case of the $B$ 
band). 
We conclude that  
there are no significant magnitude or colour variations, within the 
1$\sigma$ statistical errors, during this period. 
From all the optical photometry, we derive mean values of  
$V= 14.08 \pm 0.02$ and $V-R =0.05 \pm 0.02$ for 
Star\,1. The average blended magnitude in $B$ is $14.13 \pm 0.03$; if Star\,3 
contributes a similar fraction to the blended light as in the $V$ band, the 
correction to this value would be $\sim 0.08$~mags.

\subsection{Spectral Classification and Line Measurements}

The presence of neutral helium lines in the spectrum of Star\,1 
clearly establishes the optical counterpart as a B-type star. Furthermore, 
its optical brightness of $V \sim 14.1$ corresponds to an 
absolute visual magnitude of $\sim -4.9$ (for an SMC distance of 60\,kpc and 
E$_{\scriptsize{B-V}}$ of 0.03; see \eg Westerlund 1991). 
Thus, the implied spectral type is around B0-B1~III  
(Jaschek \& Jaschek 1987). 

The presence of the Balmer lines in emission is a clear indicator that the 
optical counterpart is a Be star (the absolute visual magnitude rules out 
the possibility that it is a supergiant, since these also show H$\alpha$ 
emission). The absorption lines of ionized Si, O and possibly 
C are entirely consistent 
with this classification, as is the probable Fe~{\sc ii} emission, which 
can occur in Be stars of spectral type B0-B5 (Jaschek \& Jaschek 1987). 
The relatively strong O{\sc ii} $\lambda$4639-42 absorption favours 
a spectral type of B1~III rather than B0~III, since this feature appears 
to be absent in the latter class (Yamashita, Nariai \& Norimoto 1977). By this 
argument, we favour the identification of the $\lambda \sim 4650$ feature 
with O{\sc ii} $\lambda$4650, rather than C{\sc iii}. A most likely 
spectral type of B1~III is thus implied. 

The equivalent widths of the prominent spectral lines were measured from the 
average spectra of Fig.~2 and Fig.~3 by summing the flux in the line after 
normalising the continuum. The results are summarised in Tab.~3. We do not 
attempt a measurement for H$\beta$ since the underlying absorption introduces 
large inaccuracies. The average value for H$\alpha$ of $15.23 \pm 0.20$~\AA\ is 
consistent with that found in 1993 Dec (Hughes 1994). 

We investigated the velocities of the strongest spectral lines through 
cross correlation with the average spectrum. No significant changes in the 
velocities of any of the lines were observed over the two nights of 
observations. The mean velocity of H$\alpha$ is $v = 155\pm2$\,km\,s$^{-1}$ 
and 
of He{\sc i}~4471 is $v = 147\pm10$\,km\,s$^{-1}$. 
These may be compared to the 
line-of-sight velocity of the SMC of $\sim 168$~km\,s$^{-1}$ (Allen 1973). 

A single Gaussian fit to the H$\alpha$ average profile yielded a width of 
$\sigma = 119 \pm 2$\,km\,s$^{-1}$ (this includes the instrumental broadening 
due to the detector resolution of $\sim$~1.3\,\AA\ or 
$\sim$~40\,km\,s$^{-1}$). 
Whilst it is true that many Be 
stars have highly rotationally broadened profiles, this is not always the case, 
particularly in binary systems (\eg Jaschek \& Jaschek 1987). The H$\beta$ 
profile is complex, hence it is difficult to obtain reliable 
measurements. However, a 
reasonable fit was achieved using two Gaussian functions to fit the narrow 
emission and broad absorption components. The central velocities and standard 
deviations of these Gaussians are $v = 132\pm2$\,km\,s$^{-1}$ and 
$\sigma = 237 \pm 33$\,km\,s$^{-1}$ for the absorption component, and 
$v = 157\pm2$\,km\,s$^{-1}$ and 
$\sigma = 68 \pm 3$\,km\,s$^{-1}$ for the emission component.

\section{Discussion}
 
The identification of RX~J0059-71 with a Be star makes this the third 
Be/X-ray binary known in the Magellanic Clouds, the other two being 
A0538-66 (\eg Charles \etal 1983) and RX~J0502.9-6626 (Schmidtke \etal 1995). 
The former system is a transient, with an orbital period of 16.7~d and 
spectral type B2~III, showing 
characteristic X-ray and optical behaviour in its active and quiescent states. 
Our optical spectrum of RX~J0059-71 more closely resembles that of A0538-66 
in its off-state (see Fig.~7 of Corbet \etal 1985), in terms of the 
absorption lines 
present, lack of He{\sc ii}~4686 emission and absence of any P~Cyg type 
profiles. However, in our spectrum of RX~J0059-71, we see H$\beta$ strongly 
in emission, whereas this line appears in absorption in 
the quiescent A0538-66 spectrum of Corbet \etal (1985). 

More importantly, the X-ray spectrum of 
RX~J0059-71 is totally unlike any of the other known Be/X-ray systems,  
the existence of a luminous supersoft component being, at present, unique to 
RX~J0059-71. Kylafis (1996) has suggested 
a model for RX~J0059-71, in which the supersoft emission arises from a torus 
of optically thick material surrounding the neutron star, with the harder 
pulsed emission resulting from direct observation of the stellar surface 
along the axis of the envelope. However, it is currently unclear why this 
system should exhibit such anomalous behaviour. In view of this fact, we have 
considered the possibility that the supersoft emission may actually arise from 
the close ($\sim 4$'') companion to RX~J0059-71 (Star\,3 in Fig.~1). We have 
only two very low quality blue spectra of this object, taken through poor 
seeing and variable cloud. No obvious spectral features 
could be distinguished above the noise; however, our optical photometry does 
suggest that this star is quite 
blue. Whilst the chance coincidence of two such unusual 
stars seems unlikely, further spectroscopic observations would be desirable
to conclusively eliminate this possibility. 

Corbet (1984) has derived a relation between the pulse and orbital periods 
of the Be/X-ray systems. If this relation applies to RX~J0059-71, an 
orbital period of $\sim 15$~d is expected (see Fig.~3 of van den Heuvel 
\& Rappaport 1987). 
However, it is not at all clear that this correlation applies in 
RX~J0059-71, if indeed this system is partially surrounded by a thick disk 
(see also A0538-66; Corbet 1986). 

\section{Conclusions}

We have identified a B1~III emission star as the probable optical 
counterpart of the 
supersoft SMC pulsar RX~J0059-71. An orbital period of $\sim 15$~d is 
predicted from the pulse period, although it is not clear that this 
calculation is applicable in this highly non-typical Be/X-ray system. 
Further optical spectroscopic and photometric observations are essential in 
order to: ({\it i}) search for a long term modulation 
($\sim$ tens of days, which is the typical orbital period range for 
the Be/X-ray systems), 
({\it ii}) investigate changes 
in the absorption/emission lines and colours, particularly in relation to the 
X-ray variability, and ({\it iii}) eliminate the possibility that the close 
companion of the Be star may actually be the source of the supersoft emission.

\section*{ACKNOWLEDGMENTS}
The data were reduced using the Starlink {\sc figaro} package and 
the {\sc ark} photometry programs. We are grateful to Keith Horne and 
Tom Marsh for the use of their {\sc pamela} and {\sc molly} routines. We 
thank the staff at the AAT and SAAO and, in particular, Darragh O'Donoghue and 
Brian Warner for the use of the UCT CCD for this project. We thank Jack Hughes 
for a communication regarding his optical data, and Ian Howarth for his 
concise comments. KAS is supported by a PPARC studentship.

\begin{table}
\caption{Journal of Spectroscopic Observations} 
\begin{flushleft}
\begin{tabular}{cccc} 
 Date & UT Start & Exp (s) & Waveband \\
  1994/11/30 & 09:53 & 600 & 6108-6863 \\
  1994/11/30 & 10:06 & " & 4364-5095 \\
  1994/11/30 & 10:19 &  " &  4364-5095 \\ 
  1994/12/01 & 09:46 &   " &   6108-6863 \\ 
  1994/12/01 & 09:59 &  " &    4364-5095 \\
  1994/12/01 & 10:10 &  " &   4364-5095 \\ 
  1996/01/19 & 11:31 & " &   4287-5052 \\
\end{tabular}
\end{flushleft}
\end{table}

\begin{table}
\caption{Optical photometry of RX~J0059-71 (Star\,1)}
\begin{flushleft}
\begin{tabular}{cccc} 
Date & $B$ (+ Star\,3) & $V$ & $R$ \\
 & ($\pm$0.04) & ($\pm$0.03) & ($\pm$0.02) \\
1996/01/24 & 14.17 & 14.06 & 14.03 \\
1996/01/25 & 14.16  &  14.11 & 14.03 \\
1996/01/26 & 14.12 & 14.09 & 14.03 \\
1996/01/27 & 14.08   & 14.11  & 14.03 \\ 
1996/01/28 & 14.13   & 14.05  & 14.02 \\
1996/01/29 & 14.12   & 14.06  & 14.04 \\
\end{tabular}
\end{flushleft}
\end{table}

\begin{table}
\caption{Equivalent width measurements of 1994 spectra}
\begin{flushleft}
\begin{tabular}{cc} 
Line & Average EW (\AA) \\
He{\sc i}~4387  & $+0.34 \pm 0.04$ \\
He{\sc i}~4471  & $+0.65 \pm 0.04$ \\
He{\sc i}~4713  & $+0.23 \pm 0.03$ \\
He{\sc i}~4921  & $+0.36 \pm 0.03$ \\
H$\alpha$ & $-15.23 \pm 0.20$ \\
\end{tabular}
\end{flushleft}
\end{table}

\begin{figure}
\caption{Finding chart for RX~J0059.2-7138 (Star\,1). 
The image was taken in the $I$-band with the SAAO 1.9\,m + UCT CCD 
and covers an area of sky $\sim 0.5' \times 0.5'$. North is up, east to the 
left. The X-ray position (J2000: 00 59 12.9, -71 38 50) 
is marked with a cross, with the short horizontal and 
vertical lines indicating the uncertainty in the declination and right 
ascension respectively. 
Differential photometry was performed relative to Star\,2. Note the close 
companion of RX~J0059-71 about 4'' to the south-east (Star\,3). There are no 
other stars within the X-ray error circle brighter then $V \sim 18$.}

\end{figure}

\begin{figure}
\caption{Average blue spectrum of RX~J0059-71 from observations taken in 1994. 
The region 
$4364-5095$\,\AA\ is covered with a resolution of $\sim 1.3$\,\AA. 
The spectrum is dominated by He{\sc i} absorption and weak lines of 
O{\sc ii} and Si{\sc iii}. 
In emission, H$\beta$ $\lambda$4861 appears very 
strongly, with broad underlying absorption. Fe{\sc ii} lines may appear in 
emission, as evidenced by the feature at $\lambda$5018.} 
\end{figure}

\begin{figure}
\caption{Average high resolution ($\sim 1.3$\AA) red spectrum. 
H$\alpha$ appears very strongly in 
emission, but we do not see evidence for any intrinsic structure. 
The absorption at $\sim 6270-6295$\,\AA\ is an atmospheric feature.}
\end{figure}


\clearpage
\newpage

\begin{thebibliography}{}

\bibitem[\protect\citename{}]{}
Allen, C. W., 1973, Astrophysical Quantities, The Athlone Press, University 
of London, 287
\bibitem[\protect\citename{}]{}
Cowley, A.P., Schmidtke, P.C., Crampton, D., Hutchings,
J.B., 1996, in van den Heuvel E.P.J., van Paradijs J., eds, Proc.\ IAU Symp.\ 
165, Compact Stars in Binaries, Kluwer, Dordrecht, p.~439
\bibitem[\protect\citename{}]{}
Charles, P.A. Booth, L., Densham, R.H., Bath, G.T., Thorstensen, J.R., 
Howarth, I.D., Willis, A.J., Skinner, G.K., Olszewski, E., 1983, MNRAS, 202, 
657
\bibitem[\protect\citename{}]{}
Corbet, R.H.D., 1984, A\&A, 141, 91 
\bibitem[\protect\citename{}]{}
Corbet, R.H.D., Mason, K.O., C\'ordova, F.A., Branduardi-Raymont, G., 
Parmar, A.N., 1985, MNRAS, 212, 565
\bibitem[\protect\citename{}]{}
Corbet, R.H.D., 1986, in Tr\"umper J.E., Lewin W.H.G., Brinkmann W., eds, 
NATO ASI Series 167, The Evolution of Galactic X-ray Binaries, 
Reidel, Dordrecht, p.~63
\bibitem[\protect\citename{}]{}
Densham, R.H.D., Charles, P.A., Menzies, J.W., van der Klis, M., van 
Paradijs, J., 1983, MNRAS, 205, 1117
\bibitem[\protect\citename{}]{}
Kahabka, P., Tr\"umper, J.E., 1996 in 
van den Heuvel E.P.J., van Paradijs J., eds, Proc.\ IAU Symp.\ 165, Compact 
Stars in Binaries, Kluwer, Dordrecht, p.~425
\bibitem[\protect\citename{}]{}
Hasinger, G., 1994, Rev.~Mod.~Astron., 7, 129
\bibitem[\protect\citename{}]{} 
Horne, K., 1986, PASP, 98, 609
\bibitem[\protect\citename{}]{}
Hughes, J.P., 1994, ApJ, 427, L25
\bibitem[\protect\citename{}]{}
Jaschek, J., Jaschek, M., 1987, The Classification of Stars, 
Cambridge Univ.\ Press, Cambridge 
\bibitem[\protect\citename{}]{}
Kylafis, N.D., 1996 in Greiner, J., ed., Supersoft X-ray Sources,  
Lecture Notes in Physics, Springer Verlag, in press
\bibitem[\protect\citename{}]{}
Long, K.S., Helfand, D.J., Grabelsky, D.A., 1981, ApJ, 248, 925
\bibitem[\protect\citename{}]{}
Parmar, A., 1994, in Holt S.S., Day C.S., eds, AIP Conf.\ Ser.\ 308, 
The Evolution of X-ray Binaries, p.~415
\bibitem[\protect\citename{}]{}
Schmidtke, P.C., Cowley, A.P., McGrath, T.K.. Anderson, A.L., 1995, PASP, 107, 
450
\bibitem[\protect\citename{}]{}
Stetson, P.~B., 1987, PASP, 99, 191
\bibitem[\protect\citename{}]{}
van den Heuvel, E.P.J., Rappaport, S., 1987, in Slettebak A., Snow T.P., eds, 
Physics of Be Stars, Cambridge Univ.\ Press, Cambridge, p.~298
\bibitem[\protect\citename{}]{}
van den Heuvel, E.P.J., Bhattacharya, D., Nomoto, K., Rappaport, S.A., 
1992, A\&A, 262, 97
\bibitem[\protect\citename{}]{}
Wang, Q., 1991, MNRAS, 252, 47P 
\bibitem[\protect\citename{}]{}
Westerlund, B.E., 1991, in Haynes R., Milne D., eds, Proc.\ 
IAU Symp.\ 148, The Magellanic Clouds, Kluwer, Dordrecht, p.~15
\bibitem[\protect\citename{}]{}
Yamashita, Y., Nariai, K., Norimoto, Y., 1977, An Atlas of Representative 
Stellar Spectra, University of Tokyo Press, 27

\end{thebibliography}
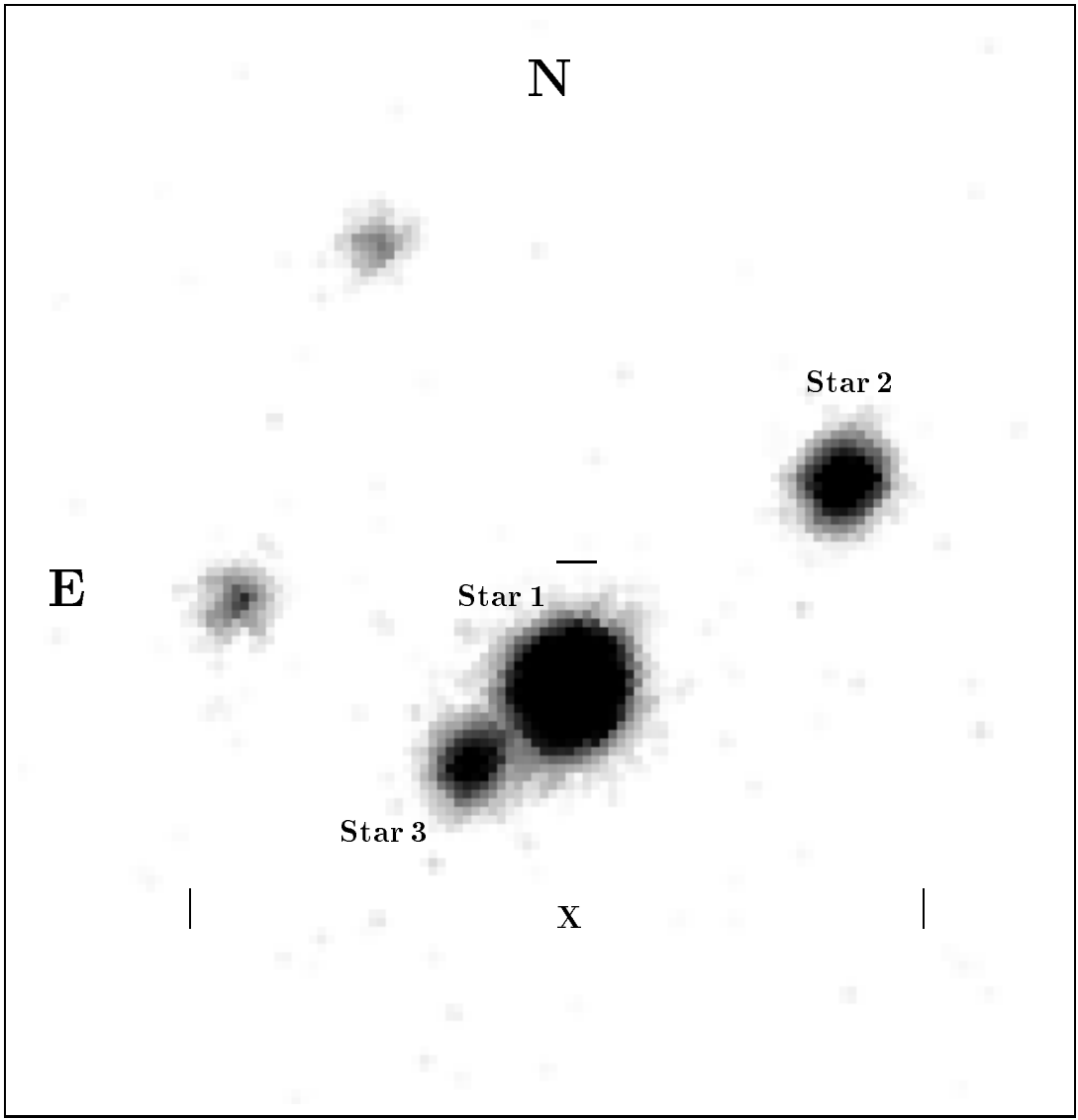
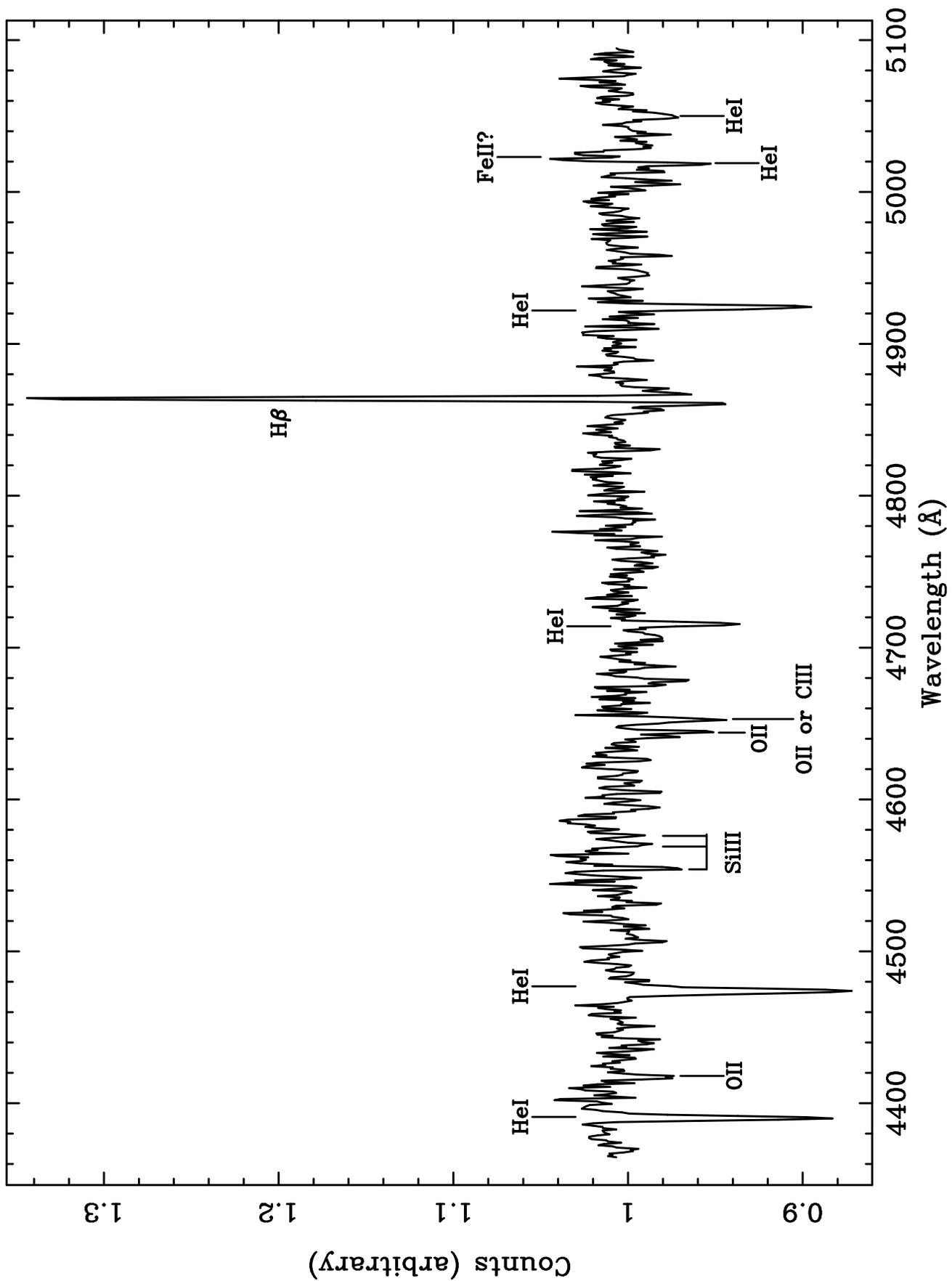
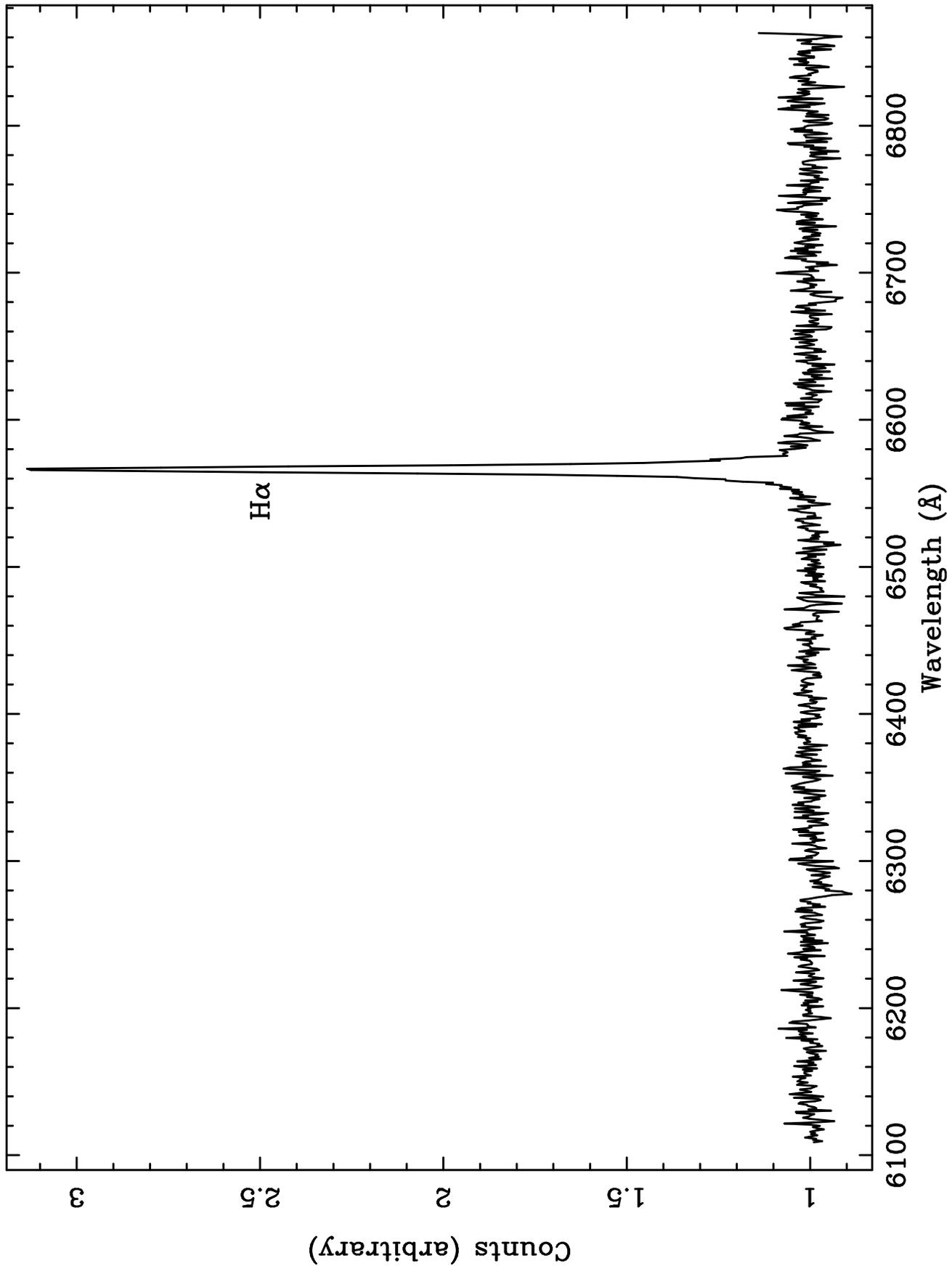
\end{document}